# Elastocapillarity-driven 2D nano-switches enable zeptoliter-scale liquid encapsulation

Nathan Ronceray[1,2], Massimo Spina[1,2], Vanessa Hui Yin Chou[2], Chwee Teck Lim[3,4,5], Andre K. Geim[6], and Slaven Garaj[1,2,3,7*]

[1] Department of Physics, National University of Singapore, Singapore 117551
[2] Centre for Advanced 2D Materials, National University of Singapore, Singapore 117542
[3] Department of Biomedical Engineering, National University of Singapore, Singapore 117583
[4] Institute for Health Innovation and Technology (iHealthtech), National University of Singapore, Singapore 119276
[5] Mechanobiology Institute, National University of Singapore, Singapore 117411, Singapore
[6] National Graphene Institute, University of Manchester, Manchester M13 9PL, United Kingdom
[7] Department of Material Science Engineering, National University of Singapore, Singapore 117575
* slaven@nus.edu.sg

Biological nanostructures change their shape and function in response to external stimuli, and significant efforts have been made to design artificial biomimicking devices operating on similar principles. In this work we demonstrate a programmable nanofluidic switch, driven by elastocapillarity, and based on nanochannels built from layered two-dimensional nanomaterials possessing atomically smooth surfaces and exceptional mechanical properties. We explore operational modes of the nanoswitch and develop a theoretical framework to explain the phenomenon. By predicting the switching-reversibility phase diagram – based on material, interfacial and wetting properties, as well as the geometry of the nanofluidic circuit – we rationally design switchable nano-capsules capable of enclosing zeptoliter volumes of liquid, as small as the volumes enclosed in viruses. The nanoswitch will find useful application as an active element in integrated nanofluidic circuitry and could be used to explore nanoconfined chemistry and biochemistry, or be incorporated into shape-programmable materials.

## Introduction

In the natural world, living structures exhibit a programmable response to external stimuli – such as mechanical, chemical, or electrical – and adapt their properties, shape, and function to match the changing environment[1]. Intelligent materials and devices[2], including emerging nanofluidic devices[3,4], are foreseen to mimic such functionalities of living matter, combining sensing, actuation, and computation. Reversible shape-switching[5–7] based on chemical and physical signals lead to design of artificial muscles[8] and microscopic robots[9]. The contemporary programmable structures[10] operate at scales above 10 µm due to mechanical limitations of solid-state thin films and elastomers that drive the switching process[8]. Here we developed a nanoscale switch operating at the 10-nanometre characteristic scale, based on nanochannels in two-dimensional materials, undergoing controlled elastocapillary transitions. We rationalized the switching process based on geometry, material, and liquid properties. This insight allowed us to rationally design the device geometries for desired functionalities and we demonstrated operation of switchable nano-containers capable of enclosing zeptoliter volumes. Our mechanically switchable nanofluidic systems could further scale down on-device chemistry beyond impressive microfluidic technologies[11–13] and could become an active element in adaptable nanofluidic circuitry.

Capillarity defines liquid-gas interfaces, drives the wetting of porous media[14] and, not least, allows trees to pull water up their xylems over tens of meters[15]. In nanometer-sized pores, capillary pressures of several hundreds of bars build up[16], and such dramatic pressures can deform the confining medium through a phenomenon known as elastocapillarity[17,18]. In semiconductor foundries, wetting and drying steps introduce a risk of elastocapillary damage and can trigger unwanted adhesion, thus hindering the miniaturization of electronic systems[19,20].

In this work, we designed 2D nanochannels in which the action of elastocapillarity is reversible, and we employ it as a mechanical switching mechanism. We used two-dimensional layered materials (graphene, hBN or $WS_2$) as building blocks, as they can achieve nanometer-sized features while displaying atomically smooth interfaces and outstanding mechanical properties[21]. The nanochannels are designed by stacking several layered 2D crystals[22] on top of the $SiO_2$/Si substrate: a) the bottom wall of the channel is defined by either a 2D crystal or base $SiO_2$ substrate; b) another 2D crystal is patterned into stripes to define the sidewalls of the nanochannel; c) and the final 2D crystal, with atomically controllable thickness, defines the top wall and operates as an actuator. This nanoscale capillarity-actuated structure – *nanoswitch* – is designed by tuning the width and height of the channels and the thickness of the top wall (**Figure 1**).

Two-dimensional nanochannels with sub-nanometer height but very stiff walls have been used to explore water[22] and ion flow[23] in the regime of extreme confinement. Ångstrom-scale 'sagging' in such stiff channels has been used to monitor the presence of water in channels[24] and explore the onset of capillary condensation[25]. These deformations were attributed to van der Waals interactions with the channel side walls, which is a distinct mechanism from the phenomenon at play in this work. Instead, we designed the 2D nanochannels to operate in different physical regime, where the channel walls could cave in under the capillary forces. To achieve the nanoswitch, the channels must satisfy two criteria: capillary forces must be strong enough to bend the top channel wall; and the adhesion between the top and bottom wall should be countered by the top layer stiffness to allow for reversible channel opening upon wetting. The interplay between elastocapillarity and adhesion determined the 10-nanometre scale motion of the top channel wall under the action of capillary pressures exceeding 100 bar, realizing new nanoscale machines which react not to traditional chemical[26] or electromechanical stimuli[27] but to liquid surface tension.

**Design and operation of the nanoswitches**

We designed nanochannels using van der Waals assembly of mechanically exfoliated crystals of graphite (Gr), hexagonal boron nitride (hBN) or tungsten disulphide ($WS_2$) transferred onto a silicon substrate with a thermal oxide layer ($SiO_2$). The graphite crystal (thickness: $h \sim 10 - 20$ nm) was patterned into stripes using electron beam lithography and reactive ion etching to define *spacers*. The channel width ($w = 400 - 1000$ nm) and the top crystal thickness ($t = 10 - 40$ nm) defined the top channel wall flexibility. For parallel channels, the pitch did not play any role in the observed phenomena, so we chose it large enough $p = 1 - 2$ μm, to allow diffraction-limited imaging of individual channels. Details on the fabrication procedure can be found in the 'Methods' section. The channel geometry is illustrated in **Figure 1**a.

The channels were filled with various liquids and subsequently the devices were blow-dried with nitrogen. The drying process could lead to the caving-in of the top wall onto the substrate, as sketched in **Figure 1**b. Similar observations at larger scales were reported in micromachined channels[19,28] and in high-aspect ratio nanopillars[29,30]. They are attributed to a twofold process: the nanostructure is deformed under the capillary pressure that builds up at the liquid-air interface[31] (**Figure 1**c), and adhesion maintains the deformation after drying.

We present in **Figure 1**d a strain map in the channel cross-section calculated for our devices with a model introduced below, leading to both compressive and tensile strain in the order of few percent. The changes between the stiff and caved-in configuration are visualized using atomic force microscopy (AFM) in tapping mode (**Figure 1**e-f), here for device D1 using a hBN top crystal, graphite spacers, and a silicon oxide substrate (abbreviated as hBN/Gr/$SiO_2$). The choice of hBN as a top wall material was motivated by its transparency, which allowed imaging the nanoswitch under a standard wide-field optical microscope. We show in **Figure 2**a-c that the mechanical changes in the crystal correspond to optical contrast (for thick enough spacers). As presented in **Supplementary Movie 1**, high-speed dark field imaging enabled capturing these mechanical deformations as they happened.

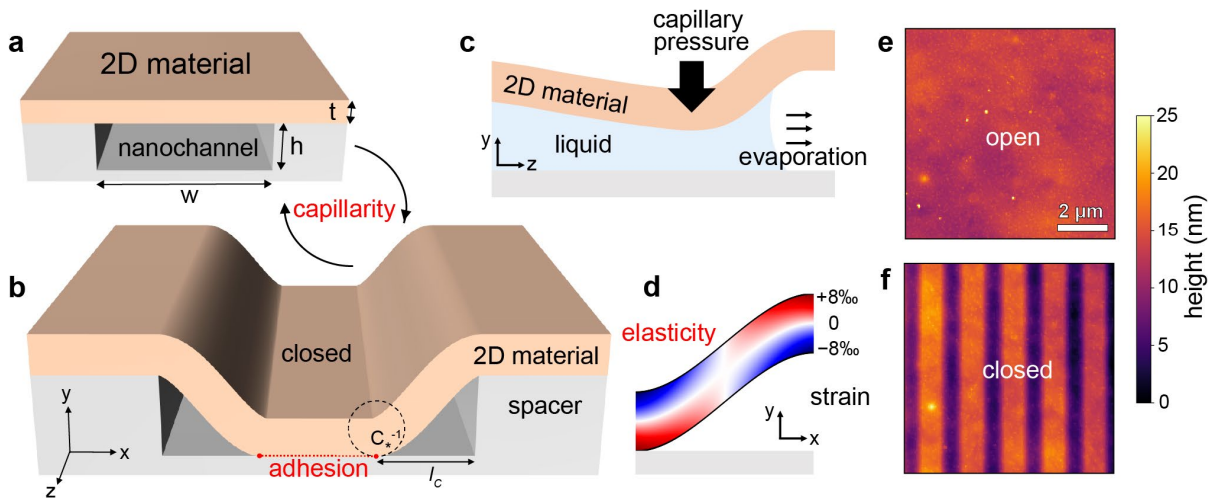

**Figure 1.** Elastocapillarity-switched adhesion in nanochannels (a) Sketch of a nanochannel device in the stiff configuration. (b) Nanochannel in its caved-in state, where the top flake adheres to the substrate. $C_\star$ is the peeling curvature parameter and $C_\star^{-1}$ thus corresponds to the radius of curvature (dashed circle). (c) Upon drying the channel, capillary pressure triggers the caving-in of the top wall (d) Calculated strain map in the bent cross-section using blue for compressive and red for tensile strain. The strain color scale refers to device D1, presented in (e,f) AFM height maps of hBN/Gr/SiO$_2$ nanochannels in their stiff and caved-in configurations, respectively

We demonstrated the operation of the switch in device D2 (**Figure 2**), with top channel walls made from hBN of thickness of $t = 34$ nm, exhibiting a good optical contrast between flat on-state and caved-in off-state. Upon critical point drying (that is, ethanol is first replaced by liquid CO$_2$, which is driven beyond its critical point), all channels are open (**Figure 2**a). Upon isopropanol (IPA) drying, we find that a portion of channels is in the off-state (dark) and the rest in on-state (bright); see **Figure 2**b. By filling the channels with water followed by drying, all channels cave in and appear dark (**Figure 2**c). The observed optical contrast can be directly correlated to the topography measured by AFM: **Figure 2**d-e shows height profiles along the colored lines in **Figure 2**a-c.

Replacing the liquid leads to different surface tension, and thus the magnitude of the capillary force during channel drying. During CO$_2$ critical point drying, there is no liquid-gas interface, and therefore no capillary pressure builds up. Water has a high surface tension $\gamma = 72$ mN/m and IPA has a lower value ($\gamma = 21$ mN/m), which is why IPA removal leaves some channels open. We used water-IPA mixtures to control the magnitude of the surface tension[32] and counted closed channels as a function of the surface tension. This defines the switch-stimulus curve in **Figure 2**f.

Closed channels remained closed for months, as long as they were dry, but they relaxed to their open position upon re-wetting. We were able to *reversibly* switch from the on- to the off-state, which could be repeated many times (**Figures S8-S9**). The only limitation that could arise from repeated drying is cumulative contaminations of the wall surfaces with impurities dissolved in the liquid, which could change the adhesion energy between the top crystal and the substrate. For this reason, we excluded acetone in our study, as it leaves residues. We could easily regenerate acetone-contaminated devices by exposing it to hot acetone baths, followed by IPA rinsing.

We identified two limiting factors for the successful elastocapillarity-induced nanoswitch. First, for the top channel wall to cave in under the capillary pressure, it must be flexible enough, which we call the *caving-in criterion*. Second, for closed channels to open upon rewetting, adhesion should not overcome wetting: this is the *reversibility criterion*. In both cases, the bending response of the top wall is crucial. Previously, the stretching and bending response of 2D materials have been studied with the so-called 'blister test' whereby a uniform pressure is applied to a 2D material suspended over a circular hole[33]. The stiffness of *multilayer* van der Waals materials was characterized recently using the same approach, yielding bending stiffnesses ranging from $10^{-14}$ to $10^{-13}$ J for crystal thicknesses in the range of 10-20 nanometers[34].

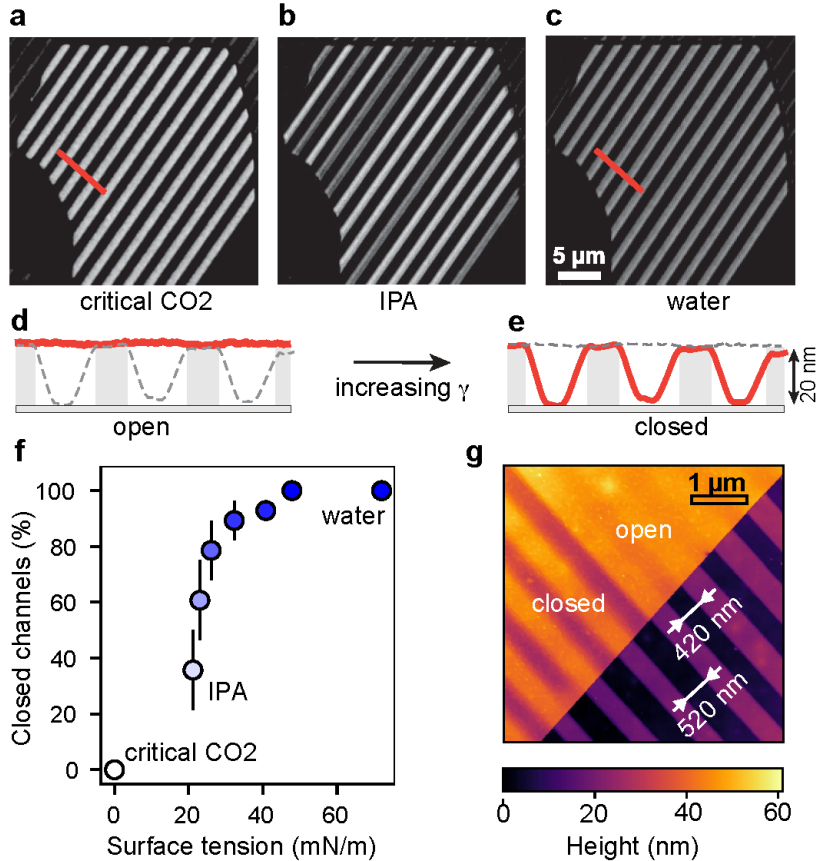

**Figure 2.** Visualizing and quantifying the nanoswitch optically and with AFM. (a-c) Optical micrographs of nanochannel device D2 after drying with different liquids. The bright and dark grey lines correspond to open or collapsed nanochannels, respectively. (d,e) Correlating the optical image to AFM profiles of channels after varying drying conditions. (f) Caved-in channels in the thick top region were optically counted as in (b) after drying with liquids of different surface tensions. IPA/water mixtures were used to interpolate the surface tension from 21 to 72 mN/m. Error bars correspond to standard deviations over 3 wetting/drying measurements. (g) Visualizing the width threshold for caving under capillary pressure: AFM map of hBN/Gr/SiO$_2$ device D3 with channels of two different widths, switched by water capillarity stimulus. The bottom-right half (dark) of the image presents open spacers, with $h = 17$ nm high graphene sidewalls. Top-left (bright) part shows the same spacers topped with $t = 22$ nm thick hBN top wall, defining the nanochannels – the upper set of channels is open, while the lower is collapsed.

**Capillarity vs. Stiffness: the caving-in criterion**

The caving-in of the nanochannel is determined by the relative magnitudes of the elastic force coming from the stiffness of the top wall, and the capillarity force. The bending stiffness $D$ of a thin film of thickness $t$ is expected to scale as $D \sim t^3$, and consequently, channels with a thin top crystal cave in more easily than those with thicker top. Wider channels should cave in more easily under a given capillary pressure, and we observed this in **Figure 2**g. Channels with $w = 420$ nm remained open upon water drying, whereas slightly wider channels with $w = 520$ nm caved in (both channels have height $h = 17$ nm and top wall thickness $t = 22$ nm).

We calculated the caving-in criterion using the instability arguments when balancing capillary pressure with the bending stiffness of the channel:[31,35]

$$\phi w^4 \mathcal{G}/Dh^2 > 1 \qquad (1)$$

Details of the calculations are given in Supplementary materials. Here we introduced the wetting surface energy $\mathcal{G} = \gamma\,(\cos\theta_S + \cos\theta_T)$ of the channel[34], given by the contact angle between the liquid and the substrate ($\theta_S$) and top wall ($\theta_T$). The prefactor $\phi$ is set by the boundary conditions[35] and reads $\phi = 1/96$. We used $\theta = 65°$ for thick 2D material-water interfaces[36], and $\theta = 0°$ for organic solvents which exhibit perfect wetting and for silicon dioxide-water interface[37]. We determined the geometrical parameters using atomic force microscopy, while bending stiffnesses were obtained from Wang *et al.*[34] and surface tension was taken from Park *et al.*[32]. The continuum medium assumption and the Young-Laplace formula are known to be valid down to few-nm confinement[16,38].

In **Figure 2**g, we see two sets of nanochannels with different widths $w_{\text{off}} = 520$ nm and $w_{\text{on}} = 420$ nm, while all the other parameters remain identical. Upon drying with water, wider channels cave in, while narrower ones remain open. Using $D = 10^{-13}$ J for 22 nm-thick hBN[34], equation (2) predicts a width threshold for caving in upon water removal of $w_{\text{th}} = 485$ nm, aligning well with the experimental results, $w_{\text{on}} < w_{\text{th}} < w_{\text{off}}$. As observed and predicted, the IPA surface tension is not high enough to trigger the caving-in for both sets of nanochannels (**Figure S4**).

**Adhesion vs. Stiffness: collapsed channels**

Accurately predicting the reversibility of caving-in – the capacity of collapsed channel to open upon wetting – is more complicated than the caving-in criterion. Interfacial properties of the adhering top and bottom walls may be affected by contaminant adsorption, chemical functionalization, and roughness of the surfaces. However, for 2D materials with smooth and reproducible surfaces, we could derive a reversibility criterion, as shown below, with sufficient accuracy to guide the design of our nanoscale switches.

To validate our method, we first derived the energy of dry, closed channels, and compared the calculated geometry of the collapsed channel with the experimental profile measured by AFM. The top crystal was modelled as an infinite plate with a bending stiffness $D$, adhering to a step-shaped substrate with adhesion energy $\Gamma$. As sketched in **Figure 1**b, due to its stiffness, the top crystal does not perfectly conform to the step and is suspended over a length $l_c$. Our model (details in Supporting Information) yields a simple polynomial shape for the bending height profiles:

$$H(x) = h\,\Lambda(x/l_c), \text{ where } \Lambda(X) \equiv 2X^3 - 3X^2 + 1. \qquad (2)$$

To experimentally test our model, we designed a slightly modified version of the switch device using WS$_2$ directly exfoliated onto a SiO$_2$ substrate patterned with photolithography, ensuring contamination-free interfaces and precise bending profiles (device D4). A top crystal, with terraced thickness ranging from $t = 6$ nm to $t = 48$ nm, was used to validate our model – presented as an AFM height map in **Figure 3**a. The corresponding slope map in **Figure 3**b is calculated as a derivative of the height map, offering evidence that the suspended length increases with increasing thickness, while the slope decreases. The experimental AFM profiles presented in **Figure 3**c are very well fitted with equation (2).

The suspended length $l_c$ minimizes the sum of the elastic energy and the adhesion energy (per unit length) $E_{\text{elas}} + E_{\text{adh}} = 12\,Dh^2/l_c^3 - \Gamma(w - 2l_c)$ and links it to the material parameters $D$ and $\Gamma$. As $l_c$ depends on the height of the step as well as the material parameters, we define a geometry-independent parameter, the *peeling curvature* $C_\star = \sqrt{2\Gamma/D} = 6h/l_c^2$. It represents the minimum curvature needed to 'peel off' the crystal from its substrate, which is obtained for $x = 0$ and $x = l_c$, as illustrated in **Figure 1**b. We can assume the adhesion energy to be thickness-independent for the thick flakes employed here[39,40], therefore, the measured peeling curvature changes only with the bending stiffness of the material.

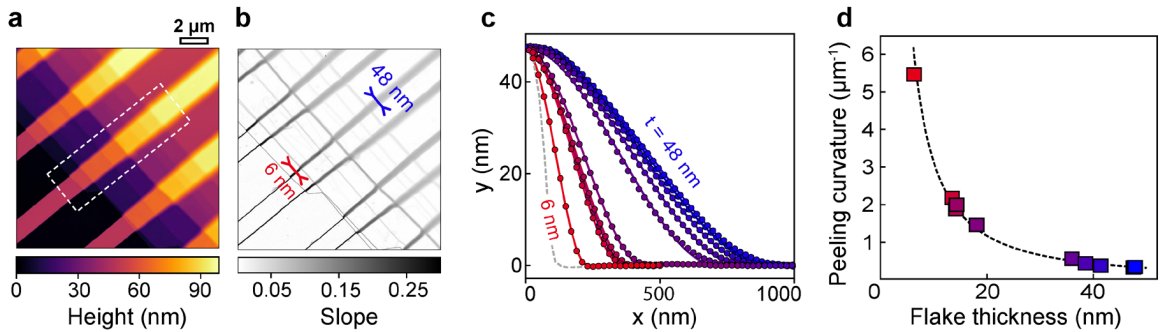

**Figure 3**. Adhesion/stiffness analysis of closed channels in device D4, obtained by exfoliating a terraced WS$_2$ crystal on SiO$_2$ spacers. (a) AFM height map of the collapsed nanochannels with varying top wall thickness $t$ ranging from 6 to 48 nm. (b) Slope extracted from the height map showing that the width of the transition zone increases with crystal thickness. (c) AFM profiles extracted along lines in the white dashed zone of (a) following the lines shown in (b). The dots are experimental points, and the solid black lines are the best fit to the polynomial profile obtained in our model. The top layer thickness was subtracted from the profiles. The x and y axis follow the convention defined in Fig. 1. (d) Peeling curvature $C_\star$ extracted from the profiles in (c), as a function of the crystal thickness. The dashed line is the best fit to a power law $C_\star \sim t^{-Q/2}$, yielding $Q = 2.88 \pm 0.08$.

**Figure 3**d depicts the experimental variation of peeling curvature with the flake thickness. The peeling curvature is measured from the AFM scans as the curvature (second derivative) at the contact line between the top and bottom walls, where the adhesion between the walls balances the bending stiffness of the top wall. The data is best fitted to a power law $C_\star \sim t^{-Q/2}$ (yielding a bending stiffness scaling $D \sim t^Q$, as shown in **Figure S3**) with $Q = 2.88 \pm 0.08$, which is close to the expected value of 3 for layered materials without slippage between planes[41], and very close to recent results using blister tests on molybdenum disulphide[34]. Our model is validated by: (a) the near-perfect fit between our model and measured AFM profiles (**Figure S1**); and (b) its quantitative agreement of with published results obtained with different experimental techniques (**Figure S2**).

Our method of stiffness/adhesion measurements has additional applications: it can be used to measure the adhesion energy between different van der Waals materials and substrates if the bending stiffness is independently determined by other techniques. Using stiffness values $D$ for different crystal thicknesses from the literature[34], we obtain $\Gamma = 82$ mJ/m$^2$ for WS$_2$-SiO$_2$ adhesion energy from the peeling curvature dependence on the thickness in the **Figure 3**d (we assumed that WS$_2$ has a similar bending stiffness as MoS$_2$). Previous studies of adhesion typically use spontaneously occurring blisters[42,43], and thus rely on assumptions on the poorly characterized content of the blisters[44]. Our method could be used for a variety of interacting materials, addressing the discrepancies in the literature regarding 2D material adhesion energies[45,46].

**Adhesion vs. Stiffness vs. Capillarity: re-wetting and the reversibility criterion**

The rewetting process starts with dry, closed channel with a strongly adhering region in the center, and two narrow side-capillaries defined by the side wall and bent portion of the top wall, and the base of width $l_c$ (see **Figure 1**b). When introducing the liquid, it creeps into the side-capillaries and enlarge their base from the dry value $l_c$ to the new value $l_{\text{wet}}$ due to the added contribution of the wetting interaction to the overall energy of the system (see **Figure S5**).

To find the new equilibrium value of the suspended length $l_{\text{wet}}$, we minimized the total energy $E_{\text{elas}} + E_{\text{adh}} + E_{\text{wet}}$, where $E_{\text{elas}} + E_{\text{adh}}$ is calculated above, and $E_{\text{wet}} = \gamma l(\cos\theta_T + \cos\theta_S)$ (see detailed derivation in Supporting Information). We achieve the full wetting when the base suspended length in the wet state becomes larger that the half width of the channel, $l_{\text{wet}} > w/2$. This is the reversibility criterion, when wetting completely releases the top wall and it corresponds to the following inequality: $\frac{w^2 C_\star}{h}\left(1 - \frac{\mathcal{G}}{\Gamma}\right)^{1/2} < 24$

**Nanoswitching phase diagram**

The elastocapillary-induced nanoswitch is achieved by choosing the right set of geometric parameters $\{w, h, t\}$ and materials $\{\Gamma, D, \mathcal{G}\}$, yielding at least 5 independent parameters. Introducing the dimensionless parameters $\alpha \equiv w^2 C_\star/h$ and $g \equiv \mathcal{G}/\Gamma$, we reduced the complexity of this system to a two-dimensional space, and the universal *nanoswitch criterion* now reads:
$$8\sqrt{3}\, g^{-1/2} < \alpha < 24\,(1-g)^{-1/2} \qquad (3)$$

where the left inequality corresponds to the caving-in criterion and the right inequality to the reversibility criterion. The value of $\alpha$ is defined by the channel geometry and the materials choice, while $g$ is defined by liquid and material interface properties.

The $\alpha - g$ phase diagram in **Figure 4**a summarizes the predictions of the nanoswitch criterion and compares them to the experimental results. The blue region corresponds to 'unbendable' channels which are too stiff to cave in under the liquid capillary pressure (for a given liquid), failing to pass the caving-in criterion. The red region corresponds to channels which fail to pass the reversibility criterion and remain closed even after rewetting. The white region passes both criteria and allows for reversible switching. Each experimental data point in **Figure 4**a represents a drying experiment followed by a check of the device configuration. White circles represent switchable channels that close upon drying and open upon re-wetting. Blue markers represent stiff channels that remain open all the time, and red markers represent channels that close irreversibly upon drying. The scale of color for the remaining markers represents the success rate for closing/irreversible collapse when multiple channels are involved, ranging from 0% (blue / red) to 100% (white), as in **Figure 2**. Near the transition line, there is a distribution of switchability of nominally equivalent channels, due to some channel-to-channel variability of the geometric and adhesion parameters. However, when moving away from this transition line, results are obtained deterministically.

Our experimental data shows good agreement with the predicted $\alpha - g$ phase diagram, demonstrating its usability in designing functional nanofluidic circuits (see below). The light red circle in the bottom-right part of the diagram only partially follows the expected irreversibility criterion, and this might be due to ill-defined interface between the bottom (SiO$_2$) and top flake (hBN), possibly affected by contaminations of surface roughness. For this configuration, only 1 out of 5 nanochannels is irreversible. However, overestimating irreversibility is not a significant issue when designing switches, since we are mostly concerned to be either in the stiff or the switchable regime.

Van der Waals nanoslits with heights set by the thickness of monolayer/bilayer graphene spacers have been used to explore a variety of nanofluidic properties[4,22,23,47]. The design constraints for such structures have been empirically established: for 0.7 nm-high, 130 nm-wide channels to remain stiff, one must use top crystals thicker than 50 nanometers[22]. Our analysis supports this observation, denoted as a star on **Figure 4**a, and those parameters are found within stiff region. Our results suggest that the geometrical criterion could be relaxed even further when performing nanofluidic experiments – even if a nanochannel is in the `switchable' region, it could readily recover its shape upon wetting, making them suitable for liquid and ion transport experiments despite caving-in in the dry state.

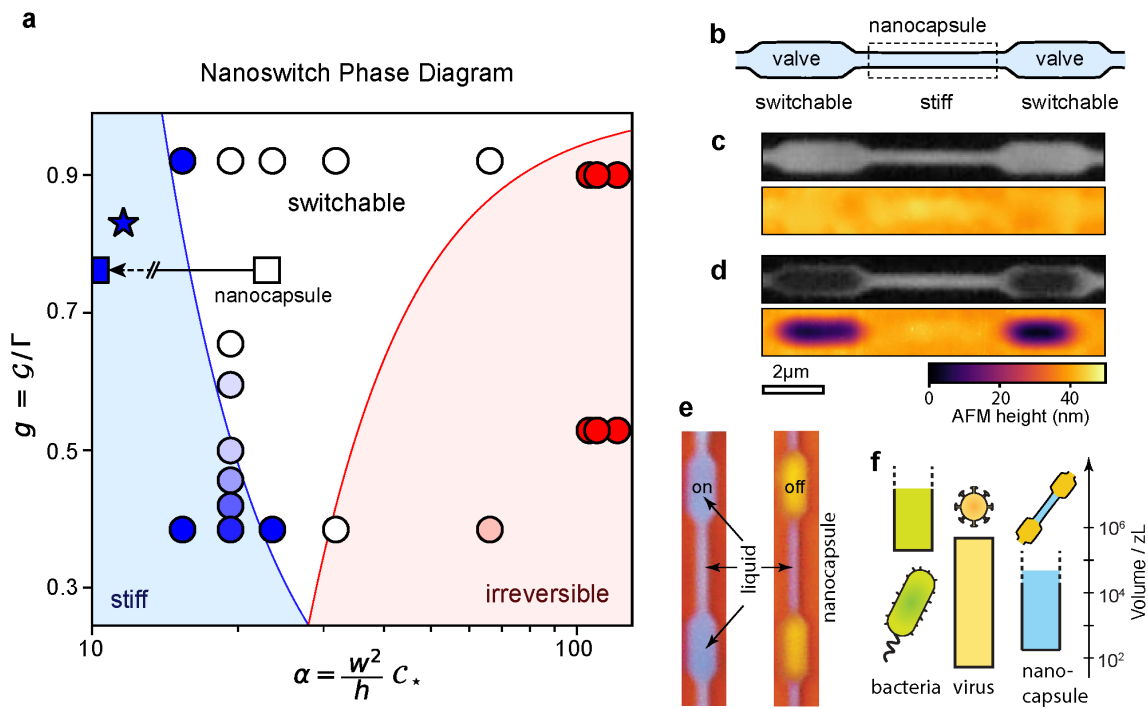

**Figure 4**. Generalization and application of the nanoswitch. (a) Universal nanoswitch phase diagram. The dimensionless parameter $\alpha = w^2 C_\star/h$ defines the channel geometry-dependent flexibility, and $g = \mathcal{G}/\Gamma$ defines the material wettability/adhesion ratio. Markers represent experiments for different geometry, solvents, and materials. Different regions in the parameter space are color-coded. For low $\alpha$, channels do not cave in upon liquid removal; this is the stiff region (blue). For high $\alpha$ and low $g$, the channels irreversibly collapse on the substrate (red). The remaining space is the switchable region, where channels undergo on/off transitions during wetting/drying (white). The star denotes the mechanical stability upper limit empirically established for atomically thin 2D nanoslits[22]. Squares denote parts of the nano-capsule: the open square is the switch part, and the blue square is for the container (off-scale, real value for container is $g \approx 4$). (b) A sketch of a self-sealing nano-capsule, consisting of a narrow stiff nanocontainer, delimited by two

switchable valves. (c) Implementation of the nano-capsule, visualized in open configuration by optical microscopy (top) and AFM (bottom). (d) The same nano-capsule in closed state, where IPA removal switches off the valves, sealing the nanocontainer with zeptoliter volume. (e) Color-scale optical image of an open nano-capsule filled with liquid (blue) with open nanoswitch gates (left). The same nano-capsule with closed off nanoswitch gates, sealing in the ~100zL of liquid inside (right). (f) Comparison of range of volumes encapsulated inside bacteria, viruses and different nano-capsules implemented in this work.

**Nano-capsule with zeptoliter volume**

The nanoswitch could be used in two ways: (1) it could react to changes in the fluid medium through its surface-tension sensitive shape response or (2) nanofluidic circuits could have switchable parts with geometry-sensitive response that could modify their structure and function. To demonstrate the latter, we used the $\alpha - g$ phase diagram to design a *nano-capsule* – an active nanofluidic system consisting of switchable gates delimiting a stiff narrower container (Fig 4b). The gates are nanochannels (width $w_G = 2\ \mu m$) that switch off due to capillarity, while the container is a narrower section of the channel (width $w_C = 600\ nm$) that does not cave in. The height of the channels is $h = 37\ nm$, the top wall is made of hBN and the bottom from graphene. The nano-capsule parameters are presented in the $\alpha - g$ phase diagram in **Figure 4**a as squares: the switch has $\alpha \approx 23$ (switchable region) and the stiff container has $\alpha \approx 4$ (out of scale).

In **Figure 4**c, the nano-capsule has open gates, as it can be seen in optical (top) and AFM (bottom) image. In **Figure 4**d, the gates were closed by elastocapillarity, deliming a container with a volume in range of ~100 zL, which could realistically be scaled further down. **Figure 4**f compares the range of enclosed liquid volumes accessible with nano-capsules with volumes of range of bacteria and viruses. A time sequence of the drying process is provided in **Figure S6**, showing the successful trapping of liquid for ~10 seconds. This trapping time, sufficient to observe fast confined phenomena, could be improved significantly through optimizing the leakage between the walls and maximizing the gate length. It is worth noting that, in case where electrolytes and solute are present, their concentration in a nano-capsule could diverge from the bulk concentration due to the drying dynamics.

Slit-shaped 2D van der Waals nanochannels are a unique platform for the study of a wide range of properties of confined fluids[3,4]. In this work, we have demonstrated the switching behavior of 2D nanochannels, driven by elastocapillarity, and developed a theoretical framework to explain the phenomenon and rationally design such nanoswitches. Switchable nanochannels should be designed in two steps: (1) choosing the top wall materials which set $\Gamma$ and $C_\star$ and (2) identifying the right spacer dimension ratio $w^2/h$.

As we employ stiff but strong van der Waals materials as nanochannel walls, we could design nanofluidic switches operating at the single-digit nanometer scale. Previously, capillarity-switched adhesion was achieved in millimeter-sized channels using electroswitching[48], but these structures rely on soft elastomers that limit their miniaturization below the micron scale.

We also designed a functional nano-capsule consisting of two switchable gates enclosing a stiff container, which allowed us to confine a zeptoliter-scale volume by using elastocapillarity to switch off the gates. These nano-capsules provide an exciting opportunity to explore phenomena at extremely small volumes, such as those found in viruses or below. As the walls of the nano-capsules are made of atomically smooth van-der-Waals materials with controllable surface chemistry, they could serve as a versatile platform for investigating nanoconfined biochemistry and chemistry[49], as well as being incorporated into shape-programmable intelligent materials or used as active elements in integrated nanofluidic circuitry.

## Methods

**Fabrication**

Patterning of spacers was done using electron beam lithography (JEOL JBX-6300FS) or photolithography (LW45B). The pattern was written on a polymer mask (PMMA and S1805 for e-beam or photolithography, respectively) and transferred to the substrate using deep reactive ion etching (Oxford Plasma Pro Cobra 100) with $O_2$ and $SF_6$ and $CHF_3$ gases for etching graphite and $SiO_2$, respectively.

**Imaging**

The optical contrast of the images, resulting from thin film interference, was a function of the spacer thickness as well as that of other layers. Rather than identifying the wavelength of optimal contrast, we recorded RGB images and used the channel with the most contrast. Optical images were acquired with an Olympus CX microscope in bright-field mode with a 100X objective using white LED illumination. For device D2, the micrographs present only the red component of the color camera, as the green and blue components were found not to contribute significantly to the optical contrast. Conversely, in device D3, the green channel had the most contrast, as shown in **Figure S7**. The selected channel of the RGB images underwent a linear contrast enhancement.
Atomic force microscopy was performed with a Bruker FastScan AFM in tapping mode. Supercritical $CO_2$ drying was performed using Leica EM CPD300.

**Wetting/drying experiments**

Channels were immersed in isopropanol, which was solvent exchanged by transferring the wet chip to a beaker containing the desired liquid. We then let channels dry under the optical microscope or by directly blow-drying nitrogen gas. In **Figure 2**a-f, closed channels were counted optically as being dark channels, between 0 and 15 in the region of interest. In cases where partial collapse was observed, a fraction corresponding to the closed length divided by the total length was assigned.

**Acknowledgements**

AKG and SG acknowledge support from Lloyds Registry Foundation. SG acknowledges support from National Research Foundation, Prime Minister's Office, Singapore, under Competitive Research Program (Award No. NRF-CRP13-2014-03), and Ministry of Education Tier 2 grant (MOE-T2EP50221-0018). NR acknowledges useful discussions with Raphaël Carpine.

**Author Contributions**

NR and SG conceived the idea. NR performed the experiments with support from MS and VHYC. NR analyzed and modelled the data. NR developed the model for the nanoswitching phase diagram, with inputs from SG. All authors discussed the results and wrote the paper.

**Competing Interests**

The authors declare no competing interests.

# Supplementary Information

**Elastocapillarity-driven 2D nano-switches enable zeptoliter-scale liquid encapsulation**

Nathan Ronceray[1,2], Massimo Spina[1,2], Vanessa Hui Yin Chou[2], Chwee Teck Lim[3,4,5], Andre K. Geim[6], and Slaven Garaj[1,2,3,7*]

[1] Department of Physics, National University of Singapore, Singapore 117551
[2] Centre for Advanced 2D Materials, National University of Singapore, Singapore 117542
[3] Department of Biomedical Engineering, National University of Singapore, Singapore 117583
[4] Institute for Health Innovation and Technology (iHealthtech), National University of Singapore, Singapore 119276
[5] Mechanobiology Institute, National University of Singapore, Singapore 117411, Singapore
[6] National Graphene Institute, University of Manchester, Manchester M13 9PL, United Kingdom
[7] Department of Material Science Engineering, National University of Singapore, Singapore 117575
*slaven@nus.edu.sg

This file includes:

**Detailed derivation of the bending profile model**
**Detailed derivation of Caving-in and Reversibility criterion**
**Figures S1-S9**
**Table S1: list of devices**
**Table S2: list of symbols**

**Detailed derivation of the bending profile model**

We use a 1D continuum mechanics model adapted from a similar system[1] for the bending profile H(x) of a thin elastic flake conforming to a step-shaped substrate. The flake is clamped on the step substrate for $x < 0$, and is suspended over a length $l_c$ at which it touches the substrate. The flake profile satisfies:
- $H''''(x) = 0$ for $0 < x < l_c$ as there is no load along the suspended part. We neglect long-range van der Waals interactions with respect to contact adhesion. Including them would lead to a correction of few % in the estimated parameters. Thus, the profile is a third-order polynomial $H(x) = a x^3 + b x^2 + c x + d$.
- The following natural boundary conditions:
  - (i)   $H(0) = h$,
  - (ii)  $H(l_c) = 0$: the flake follows the step.
  - (iii) $H'(0) = H'(l_c) = 0$: the flake is *clamped* by adhesion.

The clamping boundary condition is supported by the experimental observation of zero slope at the contact points as well as previous measurements showing a high shear stress at the 2D material-silicon oxide interface[2]. In other words, the flake cannot slip on the $SiO_2$ substrate.

For simplicity, we set $l_c$ and $h$ as unit lengths and determine the scale-free bending profile $\Lambda(X)$, defined with $X := \frac{x}{l_c}$ and $\Lambda := \frac{H}{h}$. Solving the boundary conditions system yields the coefficients of the polynomial describing the flake profile:
$$\Lambda(X) = 2X^3 - 3X^2 + 1 = (X-1)^2 (2X+1)$$
For now, we only used geometric boundary conditions. The measured value of $l_c$ minimizes the total energy $E_{\text{elas}} + E_{\text{adh}}$.

The flake elastic energy per unit length is given by $E\text{elas} = \frac{D}{2}\int_0^{l_c} C^2(x)dx$ where $C(x)$ is the local curvature of the top wall, defined by $C(x) = \frac{H''(x)}{(1+H'^2(x))^{\frac{3}{2}}} \approx H''(x)$ as the slope remains small.

**Approximation**. Here we assumed that (narrow channel approximation):
$$H'^2(x) \ll 1$$
i.e., the maximal value $H'^2(x)|_{\max} = \frac{3}{8}h\sqrt{\frac{2\Gamma}{D}} \ll 1$, which is a good assumption in our case of small channel height $h$ and larger stiffness of the top wall $D$. For our range of operational parameters this is always fulfilled, $h\sqrt{\frac{\Gamma}{D}} \sim 10^{-3} - 10^{-2}$. This assumption is further validated by the excellent comparison of experiments and theory in Fig 3c-d, Fig S1 and Fig S2.

Using the polynomial expression for the profile, we obtain $E_{\text{elas}} = \frac{6Dh^2}{l_c^3}$.

We can see in Figure S1 that all re-scaled profiles $Y = \frac{y}{h} = \Lambda(X) = \Lambda\left(\frac{x}{\ell_c}\right)$ collapse to the curve given by eq. 2, proving that we indeed derived the analytical solution (dashed line).

**Detailed derivation of Caving-in and Reversibility**

**Caving-in criterion**
The collapse of flexible channels under capillary forces has been investigated experimental and theoretically at larger scales[3,4]. We employ their calculations and revisit their phenomenological assumptions to suit our nanometer length-scales – leading to the caving-in criterion for our devices.

We introduce the deflection $\delta$ of the top wall due to the capillary pressure, and the relative deflection $\xi = \delta/h$. The deflection in the center along the width of the channel, and at the lowest point along the length of the channel will be given by balance of the strain energy and applied pressure:

$$\xi_{df} = \frac{\phi w^4}{4Dh} p_{df} \tag{S 1}$$

Where $\phi = 1/96$, $w$ is the width of the channel, $h$ is the height of the channel, and $p_{df}$ is the driving pressure that deforms the membrane, in our case the capillary pressure. The Young-Laplace formula for capillary pressure reads[5]:

$$p_{YL} = \frac{\mathcal{G}}{h(1-\xi)} \tag{S 2}$$

However, the capillary pressure is defined not necessary at the point of the highest deflection $\xi_{df}$, but it could be defined at some other deformation point $\xi_{cap} = \kappa \xi_{df}$. The $\kappa$ is a phenomenological parameter, which considers a) bending of the channel along its length from the lowest towards the dry part of the channel; b) extension of the meniscus; and c) modification of the continuum physics at the length scales below several nanometers[6,7] (the latter could be important in our system). Such effect would lead to reduction of the driving pressure by factor $\kappa$ compared to the maximum capillary pressure expected that the highest deflection point. Combining the equation (S 1) with the capillary pressure at position $\xi_{cap}$, we get expression:

$$\kappa \xi_{df}^2 - \xi_{df} + \frac{\phi \mathcal{G} w^4}{4Dh^2} = 0 \tag{S 3}$$

The above quadratic equation has no real solutions for:

$$\kappa \frac{\phi \mathcal{G} w^4}{Dh^2} > 1 \tag{S 4}$$

leading to the collapse of the channels in that case.

For the phenomenological parameter $\kappa$, van Honschoten *et al*[4] assumed the value of $\kappa = 1/2$ ("only a motivated estimate"), as it supported their experiments well. Anoop *et al*[3] followed their lead. This assumption fitted well the experimental results for their large channels.

In our case, we assume the parameter is $\kappa = 1$, due to much smaller length scales over which the meniscus develops in our devices, and the rigidity of or membranes. This value is supported by our experimental results and leads to the caving-in criterion (equation 1 in the main text).

To get a more precise estimate of $\kappa$, a numerical simulation should be performed. The resulting gain in precision might not have practical benefits, as it would be below the threshold of the device-to-device uncertainty and variability. The benefit of our Nanoswitch Phase Diagram lies in the fact that it could quickly offer good design decision when developing switching circuitry and could give insights on the influence of different parameters (solvents, materials, geometry) – without extensive simulations.

**An alternative derivation** of the caving-in criterion, leading to the same results, starts from the energy expression for channel undergoing bending of the top wall due to capillarity reads:

$$\Psi(\xi) = E_{\text{elas}} - W_{\text{cap}} = \frac{Kh^2}{2} \xi^2 + \mathcal{G} \log(1-\xi) \tag{S 5}$$

Where we introduced the effective spring constant of the channel $K = \frac{4D}{\phi w^4}$ and the wetting surface energy $\mathcal{G} = \gamma (\cos \theta_S + \cos \theta_T)$ of the channel. The channels will collapse if the energy decreases ($\frac{d\Psi}{d\xi} < 0$) and equation (S 5) does not have local energy minima in the range $0 < \xi < 1$, that is, $\frac{d\Psi}{d\xi} < 0$ in this whole range.

$$\frac{d}{d\xi}\Psi(\xi) = Kh^2\xi - \mathcal{G}\frac{1}{1-\xi} < 0 \qquad (S\ 6)$$

For the collapse channels, there should be no local minima in energy expression, hence the expression (S 6) should not have any real solutions. This leads to the requirement $\frac{4\mathcal{G}}{Kh^2} > 1$, and to the caving-in criterion in the main text (see equation 1, main text). This energy minimization requirement is equivalent to the requirement $\frac{d\Psi}{d\xi}\left(\xi = \frac{1}{2}\right) < 0$. See Figure S4 for the evolution of the energy with the relative deflection for different device parameters.

**Reversibility criterion**

The reversibility criterion can be obtained by considering an adhered, *wet* configuration (same as in Fig. 1b but with the gap under the suspended length $l$ filled with liquid, see Fig. S5) sketched here. The profile $H(x)$ is still given by the polynomial expression but it is no longer parametrized by the dry, equilibrium value $l_c$ but rather by a parameter $l$. Introduction of the liquid into the side capillary adds wetting energy to the calculations, leading to a new value of $l \geq l_c$.

The filling pathway in Figure S5 requires that the side capillary is first filled with water, which could be a comparatively slower process that drying. The value of the suspended length $l$ in wet conditions is set by the channel wetting energy $E_{wet}$ on top of the elastic energy and adhesion energy introduced in the main text. The equilibrium value of all these components is given by:

$$\frac{d}{dl}(E_{elas} + E_{adh} + E_{wet}) = 0 \qquad (S\ 7)$$

As in the main text, reasoning on the half-channel above, one obtains: $E_{elas} + E_{adh} = 6\frac{Dh^2}{l^3} - \Gamma\left(\frac{w}{2} - l\right)$, only the wetting energy remains to be computed. It is given by integrating the wetting energy $E_{wet} = -\gamma(l_{top}\cos\theta_T + l_{wet}\cos\theta_S)$, where $\theta_T$ is the contact angle between the liquid and the material of the top surface, and $\theta_S$ is the contact angle at the bottom surface, i.e. substrate. Using contour integral, we calculate:

$$l_{top} = \int_0^l \sqrt{1 + H'(x)^2}\,dx \approx \int_0^l 1 + \frac{H'(x)^2}{2}\,dx = l + \frac{1}{2}\int_0^l H'(x)^2\,dx$$

We replace $H(x)$ by its polynomial expression (equation 2, main text) and obtain:

$$l_{top} = l_{wet}\left(1 + \frac{3}{5}\left(\frac{h}{l_{wet}}\right)^2\right)$$

If we assume the same narrow channel approximation (equation S1), equivalent to $(h/l_{wet})^2 \ll 1$, we derive the wetting energy as:

$$E_{wet} = \gamma l_{wet}(\cos\theta_T + \cos\theta_S)$$

The energy balance (equation S3) must now be solved to yield an equilibrium value $l_{wet}$. Two scenarios may occur:
- If $l_{wet} < \frac{w}{2}$ wetting does not release the walls (irreversible collapse)
- If $l_{wet} > \frac{w}{2}$ wetting releases the wall.

We therefore solve the equation $l_{eq} = \frac{w}{2}$ to obtain the boundary of the soft reversibility criterion. It reads

$$-\frac{18Dh^2}{l_{wet}^4} + \Gamma - \mathcal{G} = 0$$

Therefore, we obtain:

$$\left(\frac{h}{l_{wet}}\right)^2 = h\sqrt{\frac{\Gamma(1-g)}{18\,D}}$$

which can be rephrased as:
$$\alpha = 24\,(1-g)^{-\frac{1}{2}}$$

**Figure S1**

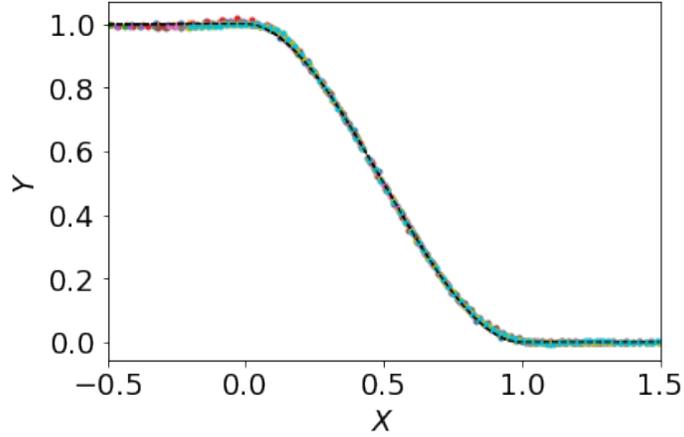

Figure S1: rescaled top wall bending profiles $Y = \frac{y}{h} = \Lambda(X) = \Lambda\left(\frac{x}{\ell_c}\right)$ for the data from Fig. 3, proving the validity of our analytical model.

**Figure S2**

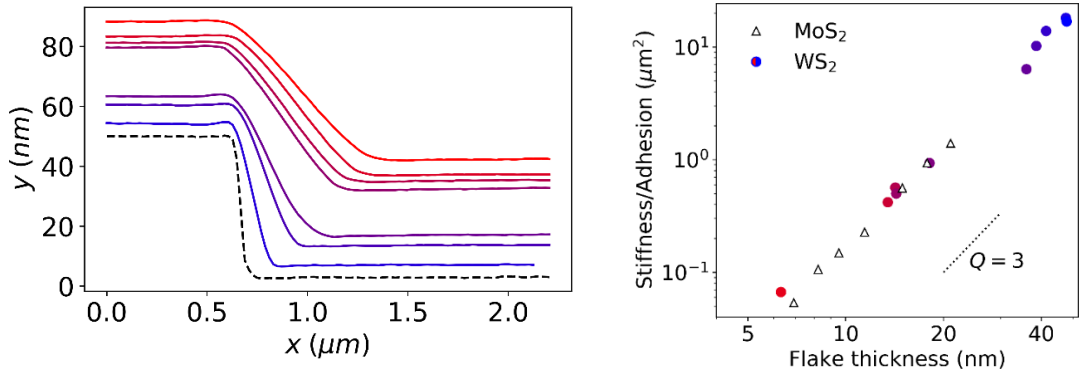

Figure S2: Left: Raw atomic force microscopy profiles without top layer thickness subtraction. Right: Stiffness/adhesion $\frac{D}{\Gamma}$ ratio as a function of the multilayer van der Waals material thickness. Colored dots correspond to data extracted from Fig. 3 in the main text, and white triangles correspond to values calculated from[8,9].

**Figure S3**

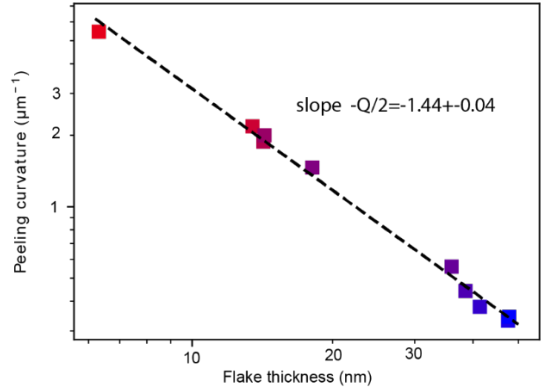

Fig. S3. Log-log plot of the relationship provided in **Figure 3**d. The dashed-line is the result of a least-square fitting to an affine function in log-log space, evidencing the power-law with an exponent very close to the expected -3/2.

**Figure S4**

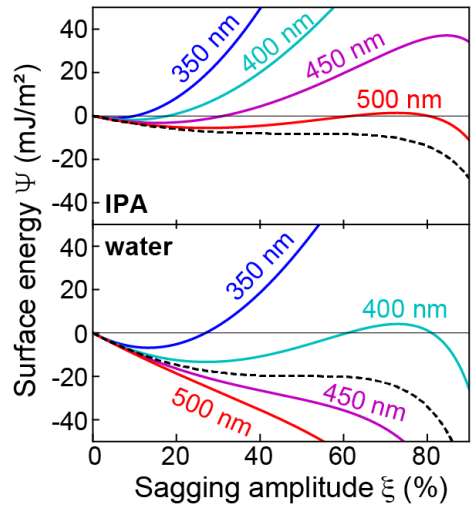

Figure S4: Predicted energy functional (per surface unit) for the capillary bending of the top wall as a function of the sagging amplitude $\xi = \delta/h$, calculated for different channel widths (labels), presented for IPA (top panel) and water (bottom panel). The dashed line predicts the width threshold separating open and closed channels for given liquids. For water, the width threshold is $w_{th} = 430$ nm, matching experimental observations.

**Figure S5**

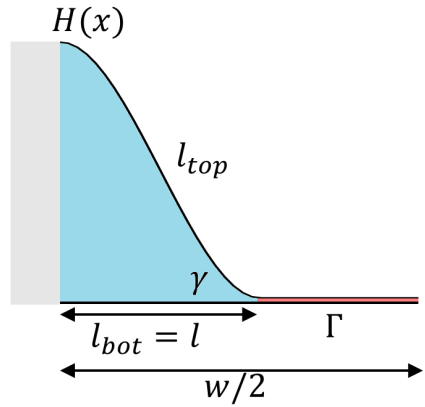

Figure S5: Sketch of the cross-section of a part of the collapsed nanochannel in the region close to the sidewall, with a narrow capillary between the sidewall and the collapsed top wall filled with liquid (blue). Adhered section of the top and the bottom wall is shown in red.

**Figure S6**

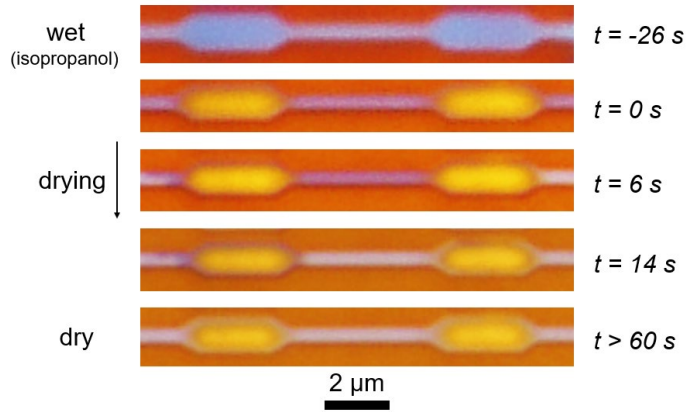

Fig S6: Liquid trapping in the nano-capsule, showing the successful trapping of liquid for ~10 seconds. The initial state (t=-26s) has the device covered with liquid. At t=0s, there is no liquid left on the device, and liquid remains trapped at t = 6s. At t=14s, the container is empty, with leftover liquid visible on the left side.

**Figure S7**

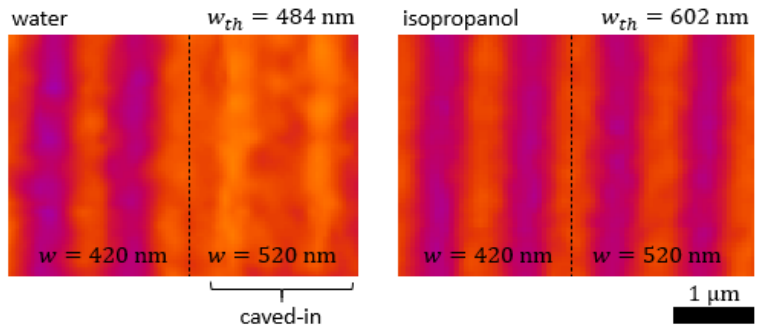

Fig. S7. Additional optical check corresponding to Fig. 2g. The green optical channel of the RGB camera was used for this device, providing suitable optical contrast. Note that in this geometry the darker parts correspond to caved-in channels.

**Figure S8**

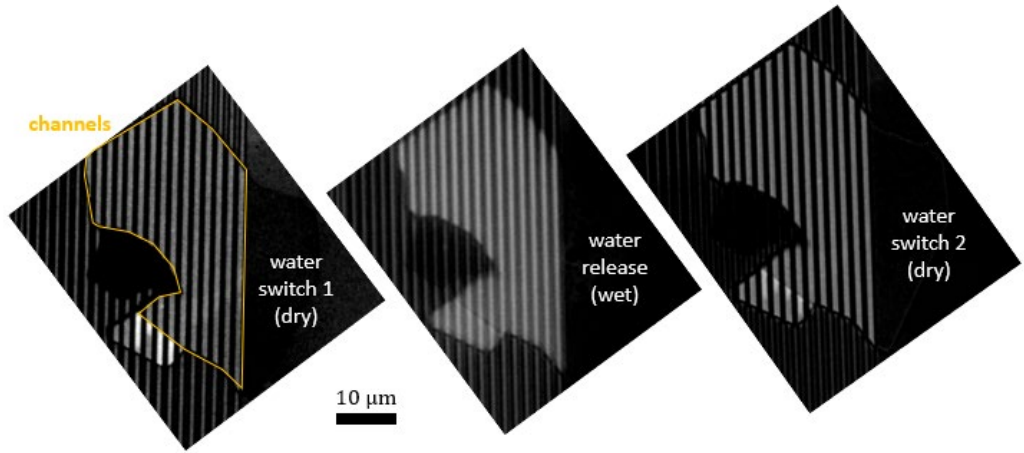

Fig. S8. Reversibility and repeatability of the switching phenomenon. The images above correspond to a full cycle of drying and wetting: we imaged device D2 after water switching (left), after which the device was imaged in water (center), and the drying was repeated (right). The bottom left region shows the expected optical contrast level for open channels in dry.

**Figure S9**

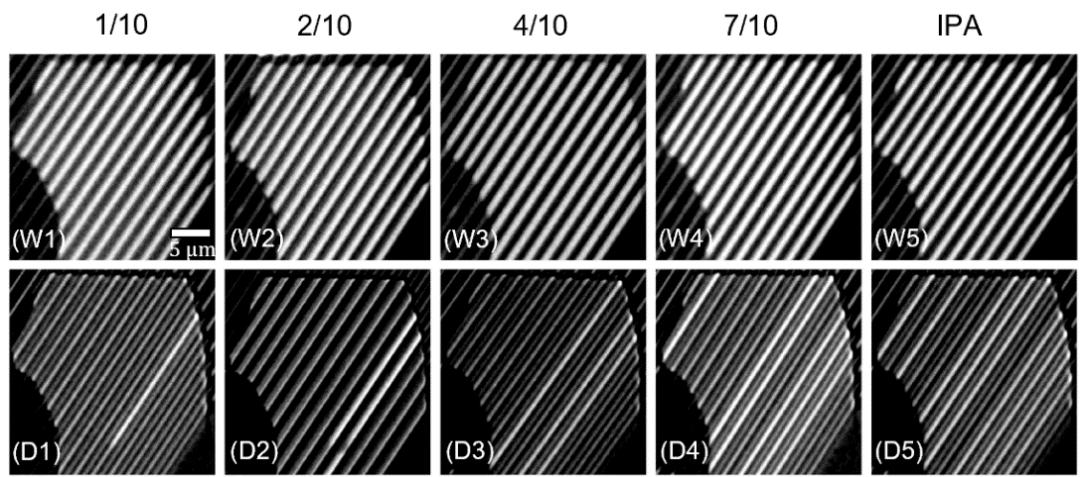

Fig. S9. Proof of the reversibility when changing the solvents (extended data from **Figure 2**). The top images, labelled (W1-5), correspond to wet channels which recovered from their previous collapse. Below, we show the outcome of the drying process with an IPA proportion corresponding to the fraction indicated above the images. This optical check confirms that channels are always released by re-wetting.

**Supplementary Movie 1**
Dark-field high-speed imaging of isopropanol removal in device D2. The sample was illuminated with white light using a mercury lamp (Nikon Intensilight C-HGFI) and the recording was performed at 500 fps (Photron FASTCAM SA3 monochrome), using a 100X air objective (NA=0.8) mounted on an upright microscope (Nikon Eclipse). Scale bar: 5 µm.

**Table S1: list of devices**

| Device number | Materials (top/spacer/substrate) | Geometry: height / width / top thickness (nm) | Comments |
|---|---|---|---|
| D1 | hBN/Gr/$SiO_2$ | 11 / 1000 / 44 | None |
| D2 | hBN/Gr/$SiO_2$ | 20 / 1100 / 12,34 | Varying top thickness, and surface tension |
| D3 | hBN/Gr/$SiO_2$ | 17 / 420,520 / 22 | Varying channel width |
| D4 | $WS_2$/$SiO_2$/$SiO_2$ | 48 / 2900 / 6-48 | Varying top thickness, irreversible collapse |
| D5 | hBN/Gr/Gr | 28/600-2000/43 | Nanocontainers |

**Table S2: list of symbols**

| Symbol | Meaning |
|---|---|
| *Geometric* | |
| $h$ | Height of the channel |

| Symbol | Meaning |
|---|---|
| t | Thickness of the top channel wall |
| w | Width of the channel |
| p | Pitch between the channels |
| x | Coordinate along the width of the channel |
| y | Coordinate along the height of the channel |
| z | Coordinate along the length of the channel |
| H(x) | Height profile of the flexible top wall, along the width of the channel |
| $\delta$ | Maximum deflection of the top wall bent by capillarity |
| *Material* | |
| D | Stiffness of the top wall |
| $\Gamma$ | Adhesion surface energy (per unit area) between top wall and the surface. |
| $E_{\text{elas}}$ | Total elastic energy of bended top wall, per unit length of the channel |
| $E_{\text{adh}}$ | Total adhesion energy between the caved-in top wall and the substrate, per unit length of the channel |
| $l_c$ | Equilibrium value of the length over which the caved-in top wall is suspended (dry state) |
| $l$ | Length over which the caved-in top wall is suspended, base of the "side-capillary" |
| $l_{\text{wet}}$ | Equilibrium value of the length over which the caved-in top wall is suspended (wet state) |
| $C_\star$ | Peeling curvature, minimum curvature needed to 'peel off' the crystal from its substrate. $C_\star = \sqrt{\frac{2\Gamma}{D}}$ |
| *Liquid and interfaces* | |
| $\gamma$ | Liquid surface tension |
| $\theta_S$ | Contact angle between liquid and the substrate (bottom surface) |
| $\theta_T$ | Contact angle between liquid and the top substrate |
| $\mathcal{G}$ | Wetting surface energy |
| $E_{\text{wet}}$ | Energy of wet caved-in channel, per unit length |
| *Parametrization* | |
| $\Lambda(X)$ | Polynomial $\Lambda(X) \equiv 2X^3 - 3X^2 + 1$ |
| $\alpha$ | Parametrization of geometric and material parameters, $\alpha \equiv w^2 C_\star / h$ |
| $g$ | Parametrization of liquid and material parameters, $g \equiv \frac{\mathcal{G}}{\Gamma}$ |

**Supplementary References**